\documentclass{ws-ijmpe}
\usepackage[super,compress]{cite}
\begin{document}

\markboth{Peter U. Sauer}{Three-Nucleon Forces}

%%%%%%%%%%%%%%%%%%%%% Publisher's Area please ignore %%%%%%%%%%%%%%%%%%%%%%%
\catchline{}{}{}{}{}
%%%%%%%%%%%%%%%%%%%%%%%%%%%%%%%%%%%%%%%%%%%%%%%%%%%%%%%%%%%%%%%%%%%%%%%%%%%%%%%%%%%%%%%%%%%%%%%%%%%%%%%%%%%%%%%%%%%%%%%%

\title{THREE-NUCLEON FORCES}

\author{\footnotesize PETER U. SAUER}

\address{Institute for Theoretical Physics, Leibniz University, Appelstrasse 2\\
D-30167 Hannover, Germany\\
sauer@itp.uni-hannover.de}

\maketitle

%\begin{history}
%\received{Day Month Year}
%\revised{Day Month Year}
%\accepted{Day Month Year}
%\comby{(xxxxxxxxxx)}
%\end{history}

\begin{abstract}
The role of three-nucleon forces in {\em ab initio} calculations of nuclear systems is investigated. The difference between {\em genuine} and {\em induced} many-nucleon forces is emphasized. {\em Induced} forces arise in the process of solving the nuclear many-body problem as technical intermediaries towards calculationally converged  results. {\em Genuine} forces make up the hamiltonian; they represent the chosen underlying dynamics. The hierarchy of contributions arising from {\em genuine} two-, three- and many-nucleon forces is discussed. Signals for the need of the inclusion of {\em genuine} three-nucleon forces are studied in nuclear systems, technically best under control, especially in three-nucleon and four-nucleon systems. {\em Genuine} three-nucleon forces are important for details in the description of some observables. Their contributions to observables are small on the scale set by two-nucleon forces. 
\end{abstract}


\ccode{PACS numbers: 21.45.+v; 21.30.-x; 24.10.-i; 24.70.+s; 25.10.+s, 25.45.De}

%\tableofcontents

\section{Historic Aspects of Nuclear Theory}

Nuclei are special many-body systems: They show shell structure, but they do not have a natural shell-forming central field which dominates the nuclear constituents, as familiar from atomic physics. On the other hand, the number of constituents is far too small to make the many-body problem describable by statistics, as familiar from condensed-matter physics. Seduced by the convincing success of atomic physics, the nuclear-physics community started to apply the same theoretical tools for the description of nuclear phenomena under corresponding assumptions: {\em Nuclei are viewed as many-body systems of rigid nucleons, driven by the dynamics of a two-nucleon force according to the rules of non-relativistic quantum mechanics.} That was the starting scenario for the microscopic nuclear theory, years back, and it still is its basis.

In the early days, the two-nucleon (2N) force was, except for its one-pion ($\pi$) exchange tail, conceptually unknown, but for a realistic description tuned to the deuteron properties and to low-energy 2N scattering data. The question of the dynamic need for an additional 3N force was outside any serious consideration, though early attempts for estimating its magnitude were undertaken \cite{brown} by courageous colleagues. In contrast to atomic physics, microscopic nuclear theory is faced with two distinct problems: First, the calculational problem: How to solve the nuclear many-body problem with sufficient accuracy? Second, the conceptual problem: Are the chosen forces appropriate for the description of nuclei? The tuning of the 2N force to 2N data and of a possible 3N force to some 3N data is a necessary condition, but not a sufficient proof  for the chosen dynamics to be realistic. 

The first phenomenological parametrisations of the 2N force were scary for theoreticians, due to a very strong short-range repulsion in its potential form; they required a particular treatment  when solving the many-N problem, i.e., they required softening the 2N potential to an {\em induced} in-medium 2N interaction. With that goal, Brueckner theory \cite{brueckner} attempts to solve the many-N problem by a special ordering of all interaction processes; it introduces the 2N reaction matrix  as the in-medium interaction, which sums up all {\em genuine} interactions between two Ns, i.e., it carries out the so-called ladder summation.  However, computations of binding energies and other nuclear observables based on that 2N reaction matrix  are just the lowest approximation on the way to a numerically converged solution of the many-N problem. Its improvement, originally discussed in the context of nuclear-matter calculations, includes the 3N-cluster \cite{bethe01,bethe02} contribution, in fact an {\em induced} 3N interaction; it is a 3N interaction, since it is irreducible in the restricted model space, chosen for calculations. Thus, from the very beginning the microscopic description of nuclear phenomena  encountered  many-N forces, forces which I call {\em induced} ones in this review. The {\em induced} forces arise in the process of practically solving the nuclear many-body problem; they are not contained in the hamiltonian which forms the dynamic basis for the theoretical description. 

Besides the arising {\em induced} forces, {\em genuine} many-N forces may have to be added to the 2N force in the nuclear hamiltonian for a conceptually complete description of the many-N problem; this is in distinction to atomic physics, whose dynamics does not need {\em genuine} many-body forces for a quantitatively successful microscopic description, though they exist in principle also there.  {\em Genuine} forces - by others often called bare or initial forces - are the interactions in which the original hamiltonian is formulated for the chosen active degrees of freedom, usually Ns. In principle, many-N forces can arise with the highest complexity of A-N forces, A being the number of Ns in the studied nuclear system. However, it is believed that the importance of many-N forces is ordered, for a numerically fully converged and conceptually complete calculation, according to the number of Ns involved in the force, and we have now good reasons for that expectation. This is why the focus of this review is on 3N forces, and it will be on the {\em genuine} 3N force, since it is for me of more fundamental interest than the {\em induced} one. The {\em genuine} 3N force, in the early days pretty uninteresting for nuclear theory, has gained more and more attention in recent years, with the advent of more and more accurate calculations for nuclear systems and of their subsequent better microscopic understanding. 

There are a number of comprehensive and instructive reviews on the subject of the {\em genuine} 3N force, some devoted exclusively to it, e.g., Refs.~\refcite{nasser} and ~\refcite{hammer}, some imbedded in reviews  \cite{epelbaum01,epelbaum02,machleidt01} on the recent advances in derivation and success of nuclear forces in general; and there is a long list of publications with the goal of hunting for 3N-force effects in data, not quoted here, but referenced well in Ref.~\refcite{nasser}. In view of that intimidating list of publications I did not feel urged to add another review. However, I was asked to formulate my opinion on the issue {\em Three-Nucleon Forces} which occupied me strongly over my whole research carrier. I admit, the review is incomplete and has a strong personal bias. I do not attempt to review techniques for deriving {\em genuine} forces. Instead, I focus on the ideas underlying the {\em genuine} 3N and many-N forces. I shall not use any equation in this review; I shall illustrate ideas by diagrams, understandable also to uninitiated readers. I shall rely on computational examples obtained by my collaborators and me, but shall also reach out to the results of others in order to emphasize similarities and differences in the obtained results. 

I arranged my thoughts as follows: The distinction between {\em genuine} and {\em induced} forces is discussed in more detail in Section 2. Section 3 describes various employed forms of {\em genuine} forces. Section 4 summarizes the existing signals for the need of a {\em genuine} 3N force in the description of nuclear observables. Section 5 gives conclusions which reflect my personal concerns on the subject. 

%%%%%%%%%%%%%%%%%%%%%%%%%%%%%%%%%%%%%%%%%%%%%%%%%%%%%%%%%%%%

\section{{\em Genuine} versus  {\em Induced}  Forces}

The {\em induced} 2N and many-N forces do not have a fundamental physics basis;  they are not measurable; they are artifacts of theoreticians arising in the process of practically solving the nuclear many-body problem, e.g.: 

\begin{itemize}
\item Brueckner theory \cite{brueckner,bethe01,bethe02} and its variants as the early shell model yield time-delayed and medium-dependent {\em induced} forces due to the strategy of selectively summing interaction processes. 
\item Instead of Brueckner theory, special unitary transformations of the underlying hamiltonian are, at present, fruitfully employed \cite{bogner01,bogner02} which squeeze the original {\em genuine} forces to act dominantly in a limited configuration domain considered most important for the nuclear phenomena under study. The driving idea of this similarity renormalization group (SRG) approach is the sound assumption that low-energy nuclear observables cannot depend on the high-momentum part of nuclear forces, i.e., on their short-range correlations. The arising {\em induced} forces remain instantaneous hermitian potentials. Even without {\em genuine} many-N forces, {\em induced} many-N forces arise in this approach. 
\end{itemize} 

\noindent
Both approaches aim to soften the original {\em genuine} forces. \\

The softening step of Brueckner theory from the 2N potential to the 2N reaction matrix deals effectively with the strong short-range correlation in the central part of the potential, but it does not radically suppress off-diagonal matrix elements between low- and high-energy momenta, as observed in Ref.~\refcite{bogner02}; that feature is a shortcoming, since it implies that 3N and higher-N cluster contributions to low-energy observables can remain sizable in Brueckner theory; furthermore, the arising medium-dependent {\em induced} forces have to be recalculated for each new many-nucleon system. 

In contrast, the SRG approach effectively decouples low- and high-energy momenta in the {\em induced} forces by construction, in a general, system-independent fashion. The {\em induced} many-N forces also appear to be a complication, but the expectation that the arising {\em induced} many-N forces become increasingly unimportant with the increasing number of involved Ns is quite plausible for low-energy observables; indeed, Ref.~\refcite{jurgenson} proved just that fact for the case of a modern, already rather soft {\em genuine} 2N force and the resulting ${^3}\rm H$ and ${^4}\rm He$ binding energies; the ${^4}\rm He$ binding contribution, arising from the {\em induced} 4N force is quantitatively negligible. The SRG-evolved hamiltonian with the {\em induced} forces is physically equivalent to the original hamiltonian with {\em genuine} forces, usually chosen with a 3N force anyhow before their SRG evolution, and it is therefore applicable as the original hamiltonian in most of the standard many-body approaches to nuclei; the exceptions are those approaches requiring local potentials, since the {\em induced} SRG potentials are highly non-local at small relative distances; attempts for creating dynamically equivalent local SRG versions \cite{bogner03} are under way. Thus, the SRG strategy of softening original {\em genuine} forces  is quite successful in speeding up the convergence of standard calculational schemes, chosen for the practical solution of the many-N problem. That fact opens the door to calculations in finite nuclei, which I have believed till now to remain intractable. \\

{\em Genuine} 2N and many-N forces are also not made by nature; they are babies of theoreticians and therefore not measurable, on similar grounds as the {\em induced} ones: Theoreticians choose the active degrees of freedom for a physically proper description of nuclear phenomena in the energy range under study; they choose them by physics experience and for practical convenience, e.g., Ns at low energies. The {\em genuine} 2N and many-N forces enter the hamiltonian and provide the conceptual basis for the dynamics. The {\em genuine} 2N and many-N forces between the constituents are therefore parts of a chosen theory form for the description of nuclear phenomena; they are formulated as instantaneous hermitian potentials, allowing the use of non-relativistic quantum mechanics as calculational framework. They are generally derived from  more fundamental assumptions on the underlying dynamics, usually a field theory, which is much harder to be cast into a tractable many-body problem. For a long time, the underlying field theory was meson theory  \cite{machleidt02} with a zoo of mesons. Chiral effective field theory ($\chi$EFT) \cite{epelbaum01,epelbaum02,machleidt01} now dominates our thinking about nuclear dynamics; $\chi$EFT is formulated in hadronic degrees of freedom, usually Ns and $\pi$s, ignoring the really fundamental degrees of quantum chromo dynamics (QCD), but  it respects the chiral-symmetry properties of QCD.

%%%%%%%%%%%%%%%%%%%%%%%%%%%%%%%%%%%%%%%%%%%%%%%%%%%%%%%%%%%%

\section{{\em Genuine} Forces for the Description of Nuclear Phenomena}

 {\em Genuine} 2N and possibly many-N forces are the dynamic basis for a microscopic and realistic description of nuclear phenomena, usually called an {\em ab initio} description. Microscopic description means the derivation of nuclear properties from the interaction between the chosen active constituents of the nuclear many-body system. Realistic description means that the underlying 2N force describes  the isolated 2N system with high precision, i.e., the deuteron and the 2N scattering data in the considered energy regime. A microscopic and realistic description is based on an educated guess on the physically important degrees of freedom, chosen for the Hilbert space of calculations, a guess usually tested already before to be sound, and on a usually non-unique derivation of their forces. It is a model, therefore non-fundamental and not accessible to experimental validation; it is developed by theoreticians for the description of nuclear phenomena in the particular energy range under study and for ease of numerical calculations. 
 
In the energy range up to $\pi$-production threshold the chosen active degrees of freedom are Ns with properties as observed for free Ns; instantaneous forces between the Ns and electroweak currents of the Ns are developed for a many-N hamiltonian which describes the dynamics within the framework of non-relativistic quantum mechanics. 

In the energy regime above the $\pi$-production threshold the $\pi$ has to be introduced as an additional active degree of freedom to the Hilbert space. Instantaneous forces between and electroweak currents of all active constituents are developed for a many-body hamiltonian which describes the dynamics within the framework of non-covariant quantum mechanics; the kinematics for the $\pi$ has to be relativistic, whereas the kinematics for the Ns can remain non-relativistic. 

The explicit potential forms of 2N and many-N forces are different in the two distinct choices for the theoretical description, though their physics effects on definite observables have to be the same. The derivation of the forces is usually from an underlying field theory covering the active constituents and some additional hadronic degrees of freedom. The steps to corresponding and in some sense equivalent instantaneous potentials involve the freezing of degrees of freedom and are therefore non-unique. Fig.~1 illustrates that process of freezing degrees of freedom by the example, when meson theory is chosen as underlying field theory. When $\chi$EFT is the underlying field theory, a corresponding freezing process has to be done towards instantaneous potentials, only the field-theoretic building blocks are different, e.g., the only mesons are $\pi$s, baryon vertices with more than one $\pi$, vertices between $\pi$s and contact vertices between baryons contribute; furthermore, the intermediary freezing step to baryon potentials involving $\Delta$ isobars is at present not done in the $\chi$EFT approach. The meson degrees of freedom appear in their exchange between Ns; they are frozen by eliminating the energy dependence and the subsequent time delay of the exchanges as indicated in Fig.~1; the freezing is often technically achieved by identifying the on-shell Feynman-diagram processes of the field theory with the corresponding potential description of quantum mechanics and interpreting the equality also off-shell. The wide spectrum of alternative freezing procedures to instantaneous potentials is summarized in Ref.~\refcite{phillips01}; a popular and quite successful procedure is by unitary transformations which isolate and eliminate the meson degrees of freedom; it was developed in Refs.~\refcite{fukuda} and \refcite{okubo}, and it is now successfully used in the $\chi$EFT approach \cite{epelbaum03} to nuclear potentials. 

\begin{figure}[th] 
\centerline{\includegraphics[width=0.9\textwidth]{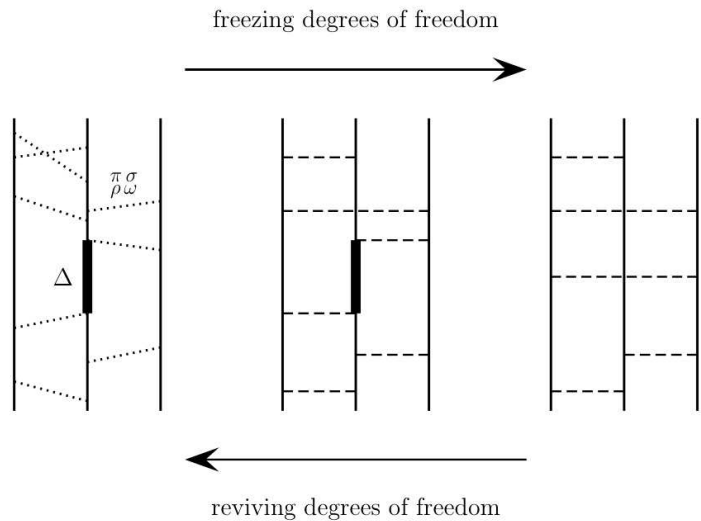}}
\caption{\footnotesize Example for the process of freezing and reviving degrees of freedom contained in a field-theoretic description. Meson theory with $\pi$, sigma, rho and omega mesons is chosen as example. Dotted slanted lines indicate the time-delayed meson exchanges, dashed horizontal lines instantaneous potentials. Reading the diagrams from left to right, firstly meson degrees of freedom are frozen into instantaneous potentials between baryons, secondly also isobar degrees of freedom; in the end 2N and 3N forces arise as instantaneous potentials. Reading the diagrams from right to left, degrees of freedom are revived, resolving the instantaneous potentials into simpler exchanges; in the first step the isobar degrees of freedom, here the $\Delta$ isobar, are revived as in Subsection 3.2, e.g., in a part of the 3N potential; the baryon interactions remain instantaneous potentials after this first revival step.}
\end{figure}

How can the potentials describing the {\em genuine} 3N force be parametrized efficiently? Usually the various contributions to the 3N force are differentiated by the distinct topologies of the Feynman diagrams in the irreducible field-theoretic processes from which the instantaneous potentials are obtained; that visualization is illustrative, but quantitatively not directly instructive. Instead, one would like to have a minimal, but complete set of independent isospin-spin-momentum (or coordinate) generators for the most general description of the 3N forces, each weighed with scalar functions of the two internal momenta (or coordinates) prior and after the interaction processes; the scalar functions could then be discussed in detail. Of course, one has the transparent description of 2N potentials by the central, spin-spin, tensor, spin-orbit and quadratic spin-orbit terms in isospin singlet and isospin triplet channels in mind; but this description is not the most general one, it is based on a local approximation of the full field-theoretic expressions, even if the augmenting scalar functions of the potentials were nonlocal. Ref.~\refcite{pask}  started the corresponding game of deriving linearly independent isospin-spin generating terms for the 3N potentials, Refs.~ \refcite{krebs01} and \refcite{phillips02} give sets in local approximation. However, even in local approximation the minimal number of necessary generators is discouragingly large for a general discussion of the 3N force; nevertheless, Refs.~\refcite{krebs01} and \refcite{pisa03} gained some important insight this way in the difference between suggested 3N forces and in their various dynamic contributions. In contrast to the attempts of illustrating, describing and understanding possible {\em genuine} 3N forces and in an act of desired practicability, the consumers of the forces now seem to demand the matrix elements directly in a 3-body partial-wave basis for a convenient digestion by their computational machineries employed in solving nuclear many-body problems.

%%%%%%%%%%%%%%%%%%%%%%%%%%%%%%%%%%%%%%%%%%%%%%%%%%%%%%%%%%%%

\subsection{Description of nuclear phenomena at low energies} 
 
This description is for nuclear phenomena at energies below the $\pi$-production threshold. The only active degrees of freedom are Ns. For a better comparison with the more complex situation of higher energies in the next subsection, the purely nucleonic Hilbert space and the forces between the Ns are shown diagrammatically in Fig.~2. 

\begin{figure}[th]
\centerline{\includegraphics[width=0.9\textwidth]{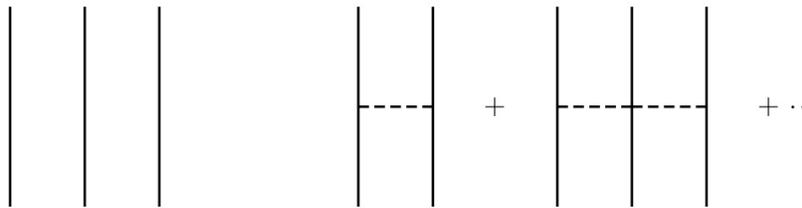}}
\caption{\footnotesize Purely nucleonic Hilbert space for the description of nuclear phenomena at low energies (left), the shown example being for N number 3, and its {\em genuine} 2N and 3N forces (right). The inclusion of 2N and 3N forces are at present standard; the ellipsis stands for the possibility of more complex many-N forces in a Hilbert space of N number larger than 3.}
\end{figure}

In the early days, the {\em genuine} 2N force was, in its tail behavior,  derived from the field-theoretic single-$\pi$ exchange; in contrast, the inner part of the interaction was chosen phenomenologically, usually as a local potential for numerical convenience. Later on, the 2N force was derived as one-boson exchange from a meson field theory \cite{machleidt02} with a zoo of mesons. Arising {\em genuine} 3N forces were often substituted by phenomenological ones, the phenomenology being enriched \cite{urbana,illinois} with some conceptual insight. The early attempts for a conceptual consistency between 2N and 3N forces were done on the level of 2$\pi$ exchange and resulted in the Paris 2N potential \cite{paris} and the Tucson-Melbourne \cite{tucson01} and Brazil \cite{brazil} 3N forces; the Tucson-Melbourne 3N force was updated in Ref.~\refcite{tucson02} for a flaw of chiral symmetry in its original version.

Presently, the most-advanced derivation of nuclear forces is based  \cite{epelbaum01,epelbaum02,machleidt01} on $\chi$EFT. $\chi$EFT provides single- and multiple-$\pi$ exchanges for the description of the long- and intermediate-range interactions and contact interactions between Ns for the description of the short-range correlation between the Ns; the description is ordered according to small external momenta $Q$ or the $\pi$ mass as  low-energy expansion parameter, the expansion order being powers of that low-energy parameter over a hard scale, the chiral-symmetry breaking scale  $\Lambda$ of about 1 GeV, i.e., $(Q / \Lambda){^{(n+1)}}$, and phrased $N{^n}LO$, $n$ being the degree next ($N$) to the leading order $LO$.  The $\chi$EFT description is designed for low-energy use and is therefore constructed to be especially soft with a decreased extension to higher momenta. The $\chi$EFT approach yields consistent {\em genuine} 2N, 3N and many-N forces. The 2N forces presently in use are given in Refs.~\refcite{machleidt01a} and \refcite{epelbaum03a} and are determined up to the chiral order $N{^3}LO$. The corresponding 3N force, mostly employed, is determined \cite{epelbaum03b} only up to $N{^2}LO$; an application-friendly local approximation is developed in Ref.~\refcite{navratil01}. The derivation of the 3N force, consistent with the 2N force of the presently desired chiral order $N{^3}LO$, is close \cite{bernard} to its completion, but not yet available in full for applications. The corresponding 4N force arises first  in chiral order $N{^3}LO$ and is given in Ref.~\refcite{epelbaum03c}.

The derivation of nuclear forces by $\chi$EFT  is at present without alternative. It is deeply rooted in symmetry principles of QCD, but it is also not fundamental. The derivation of instantaneous potentials requires a number of turning points which have to be passed in a non-unique way: The underlying relativistic field theory can be formulated without and with $\pi$s, solely with Ns  or with additional baryons, though the $\pi$-less theory does not appear useful in the present context; the theory with additional baryons besides Ns will be discussed later. Power counting is an art with non-unique options. The step to instantaneous potentials, applicable in quantum mechanics, can be carried out by different procedures. $\pi$ loops are regularized by dimensional or by spectral-function regularization. Remaining external high momenta have to be cut off. Relativistic corrections arise, but can be dealt with in different ways. The employed 2N scattering equation is not always consistent with the many-body equations of strict non-relativistic quantum mechanics, used subsequently in applications of the forces. I do not consider the various turning points for the construction of potentials as a flaw of the theory; if carried out consistently in all parts of the 2N and many-N potentials, the non-unique choices are absolutely acceptable. But those choices prove that the resulting 2N and 3N potentials are not measurable quantities; they have to be considered as parts of another special, though well-founded dynamic model for the nuclear forces.

As all 2N potentials in order to become realistic, the $\chi$EFT 2N force has to be tuned to the low-energy 2N data. Tunable are the parameters related to the contact terms in the potential; there are 26 parameters at the chiral order $N{^3}LO$, 24 are charge-independent, 2 charge-dependent. Ref.~\refcite{machleidt01a} also fits 3 low-energy constants for the $\pi$ exchanges; it uses dimensional regularization for the  $\pi$ loops; the fit is energetically carried up to the $\pi$-production threshold. The 2N tuning of Ref.~\refcite{epelbaum03a} has not been pushed yet up to the $\pi$-production threshold; it considers all low-energy constants for the $\pi$ exchange as given by $\pi$N physics; it uses spectral-function regularization for the $\pi$ loops, requiring a cut-off mass for that regularization. Both fits  introduce regulator functions for the external momenta of the potentials, necessary for their subsequent applications requiring their iterations; the  cut-off mass and the functional form of the regulator function are chosen by intuition and on conceptual grounds; the resulting momentum cut-off is extremely sharp; regulator mass and functional form are not to be considered as  part of the set of tuning parameters, though they strongly influence the practical fits. Ref.~\refcite{machleidt01a} suggests a single potential form, fitted with high precision. Instead, Ref.~\refcite{epelbaum03a} uses a range of regularization parameters and thereby provides families of $\chi$EFT potentials of comparable tuning quality for each expansion order; the spread of obtained many-N results reflects the remaining uncertainty of the determined potentials due to choices in tuning. The conceptual uncertainty due to the finite order, practically reached in the chiral expansion, remains inherent in both tuned 2N potentials. 

The 2N tuning determines the 2N potential, but also largely the 3N force; this fact yields the beautiful consistency between 2N and many-N forces in the $\chi$EFT approach. However, 2N tuning does not completely determine the accompanying  3N force; the 3N contributions arising from the one-$\pi$-exchange-contact and from the purely-contact topologies of chiral order  $N{^2}LO$ carry one tunable parameter each, the contributions of chiral order $N{^3}LO$ are then completely determined. Thus, the 3N force also has to receive tuning, usually done with observables of the 3N and 4N systems, e.g., the ${^3}\rm H$ and/or ${^4}\rm He$ binding and/or the neutron-deuteron (nd) scattering lengths and/or ${^3}\rm H$ $\beta$-decay and/or the ${^4}\rm He$ charge (point-charge) radius; correlations \cite{pisa03} among some observables in that set of tuning possibilities makes a precise determination of the two tunable 3N-force parameters difficult;  Ref.~\refcite{navratil01} carries out a fit of the 3N force of chiral order $N{^2}LO$ in its local version; its parameters are chosen to be consistent with the 2N potential of Ref.~\refcite{machleidt01a}, though the regulator for external momenta is not. The $\chi$EFT 3N and 4N forces of chiral order $N{^3}LO$ are completely determined, once the corresponding 2N force and the 3N-force part up to $N{^2}LO$ are fixed; they should always be employed only with its matching 2N force.

As the SRG approach \cite{bogner01,bogner02} to {\em induced} forces, discussed in Section 2, and its renormalization group (RG) origin \cite{bogner02} with its potentials $\rm V_{\it low \, k}$, $\chi$EFT is driven by the idea that low-energy nuclear physics can depend only on the low-momentum part of nuclear forces. In contrast to the SRG and RG approaches, $\chi$EFT implements that idea already at the level of the underlying field theory from which the quantum-mechanical potentials are then derived. SRG-evolved $\chi$EFT potentials  are purged from higher-momentum unessentials which remained despite the high-momentum cut-offs in the original $\chi$EFT potentials; they provide the practical advantage of faster converging computations with and thereby allow applications of  $\chi$EFT forces in rather complex nuclei.

%%%%%%%%%%%%%%%%%%%%%%%%%%%%%%%%%%%%%%%%%%%%%%%%%%%%%%%%%%%%

\subsection{Unified description of nuclear phenomena at low and intermediate energies} 

In the energy range up to about 0.5 GeV internal energy, $1\pi$-, $2\pi$- and $3\pi$-production channels are open, though single-$\pi$ production dominates \cite{sauer01} the inelasticity; it occurs mostly in 2N isospin-triplet scattering, i.e., via the mechanism of the virtual excitation of one of the Ns  to a $\Delta$ isobar, the latter being coupled to resonating $\pi \rm N$ states. In this extended energy regime it is therefore conceptually advisable to introduce the $\pi$ and the $\Delta$ isobar as additional active degrees of freedom in the Hilbert space besides the N. 

An attempt for such a unified description was done in Refs.~\refcite{sauer01} and \refcite{poepping} and by others, e.g., in Ref.~\refcite{lee}; it was done at the time of $\pi$ factories, when the physics community hoped to get new insight into nuclear forces by studying reactions with free $\pi$s. The chosen Hilbert space is built from the two types of baryons, i.e., the N and the $\Delta$ isobar; it consists of a nuclear sector and two additional sectors, one in which a single N is replaced by a $\Delta$ isobar and one in which a $\pi$ is added to the configuration of Ns. The Hilbert space is shown in Fig.~3. The hamiltonian is restricted by {\em fiat}, not to produce more complex configurations, i.e., channels  with more than one $\Delta$ isobar or channels with more than one $\pi$ or channels  with a $\Delta$ isobar and with $\pi$s. 

Under those channel constraints, any hermitian hamiltonian yields a fully unitary theory, i.e., the multi-channel S-matrix connecting all physical channels without $\pi$s and with a single $\pi$ is unitary. The standard scattering theory had to be extended \cite{poepping} to also include particle production and annihilation. The range of possibilities for such a hamiltonian is wide; Refs. \refcite{sauer01} and \refcite{poepping} chose the following simple form illustrated in Fig.~4: It contains two-baryon potentials derived from all possible meson exchanges according to the rules of standard meson theory; this type of derivation was without alternative when this theoretical frame work was constructed; the two-baryon potentials couple purely nucleonic channels with each other, as expected from a 2N force, but they also couple them with $\rm N \Delta$ channels. In many dynamic models the coupling to channels with two $\Delta$s appears strong; in the present context those contributions are not explicit in the hamiltonian, they are omitted due to the omission of 2$\Delta$ and 2$\pi$ channels on the ground that 2$\pi$ production is weak at the energies to be considered; any possibly existing  2$\Delta$ dynamics is hidden in the instantaneous two-baryon potentials. The hamiltonian also contains a one-baryon piece, the $\pi \rm N \Delta$ vertex, mediating $\pi$ production and absorption and mediating  $\pi \rm N$ scattering in the ${\rm P}_{33}$ partial wave with its resonance to which the one-baryon vertex is tuned; a $\pi \rm N$ potential is to be added for the non-resonant $\pi \rm N$ partial waves, not shown in Fig.~4. This hamiltonian has a particular characteristics \cite{poepping} for the $\Delta$ isobar, welcome for the described physics: The $\Delta$ isobar is a bare baryon, it cannot be produced experimentally; the corresponding S-matrix element for a possible experimental production is exactly zero; observable are only the $\pi \rm N$ states, coupled to the $\Delta$ isobar and resonating in the ${\rm P}_{33}$ partial wave. In contrast, the N is assumed to be already dressed, a single-N vertex, providing an additional $\pi$-production mechanism, is not taken into account. 

According to Ref.~\refcite{sauer01} and to references therein, calculations were carried out for most aspects of that ambitious hamiltonian, e.g., for all reactions in the two-baryon sector, i.e., $ \rm NN \rightarrow NN$, $ \rm NN \rightarrow d \pi$, $ \rm NN \rightarrow NN \pi$, $ \rm d \pi \rightarrow d \pi$,  $ \rm d \pi \rightarrow NN \pi$ and $ \rm d \pi \rightarrow NN$  up to 0.5 GeV c.m. energy. The hamiltonian should have been tuned to the data of all those reactions with baryon number 2; but, in practice, its complete version was not even tuned well to the elastic low-energy 2N data, and it is therefore judged not to be reliable enough for use in few-N physics at low energies, relevant for the discussion in this review. 

\begin{figure}[th]
\centerline{\includegraphics[width=0.8\textwidth]{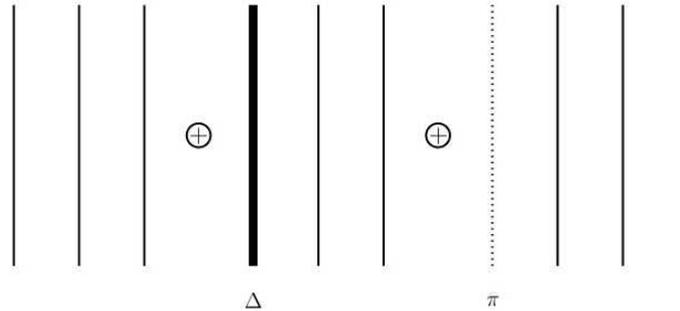}}
\caption{\footnotesize Hilbert space for the description of nuclear phenomena at low and intermediate energies. The shown example is for baryon number 3. Compared with the purely nucleonic Hilbert space of Fig.~2, it is extended by two sectors, in which one N is replaced by a $\Delta$ isobar and one $\pi$ is added to the Ns. $\pi  \rm N$ scattering is described in the corresponding Hilbert space of baryon number 1. The 2N reactions without $\pi$s and with a single $\pi$ are described in the corresponding Hilbert space of baryon number 2.}
\end{figure}

\begin{figure}[th]
\centerline{\includegraphics[width=0.8\textwidth]{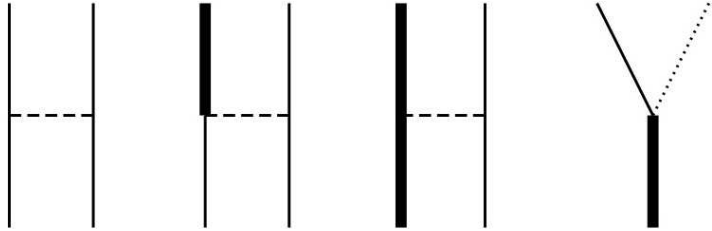}}
\caption{\footnotesize Hamiltonian describing the nuclear dynamics in the Hilbert space of Fig.~3. The interactions are of two-baryon nature, coupling purely nucleonic channels with those containing a $\Delta$ isobar; the latter ones are coupled to the pionic channels by the single-baryon vertex. The hermitian conjugate processes are to be added.}
\end{figure}

However, the explicit appearance of the $\Delta$ isobar in the Hilbert space has one particular conceptual charm, most important for the physics of this review: The same hamiltonian is able to describe intermediate-range $\pi$ physics and the low-energy nuclear dynamics with {\em genuine} 2N and many-N forces, simultaneously and consistently. E.g., it yields  3N, 4N and many-N forces by iteration of two-baryon potentials; they are irreducible in the purely nucleonic Hilbert sector, but are resolved into their two-baryon pieces in the extended Hilbert space of Fig.~3; they are therefore not instantaneous potentials as the {\em genuine} forces of Subsection 3.1, but they are time-delayed, reducible into simpler dynamic building blocks, and they become medium-dependent. In standard meson theory and in $\chi$EFT, instantaneous 2N, 3N and many-N potentials arise from freezing {\em all} non-nucleonic degrees of freedom, as illustrated in Fig.~1; but vice versa, as done in this approach and also illustrated in Fig.~1, the important $\Delta$-mediated contributions to {\em genuine} 3N and many-N forces are resolved into iteration of two-baryon potentials, when the $\Delta$-isobar degree of freedom is kept explicitly. Those contributions arise prior to and independent of any calculation; this is why they are to be classified as {\em genuine} forces. The occurrence of  important many-N forces by $\Delta$ mediation was the reason why additional irreducible three-baryon forces have not been considered yet for the hamiltonian of Fig.~4.

The $\Delta$-mediated many-N forces come together with a 2N force which also receives, besides its instantaneous part,  time-delayed and medium-dependent contributions as illustrated in Fig.~5. The time-delayed and medium-dependent part of the 2N force gives rise to a weakening of its attraction in the medium as compared to the free 2N interaction, when other nucleons simultaneously propagate; this is the so-called 2N {\em dispersive effect}. 

\begin{figure}[th]
\centerline{\includegraphics[width=0.9\textwidth]{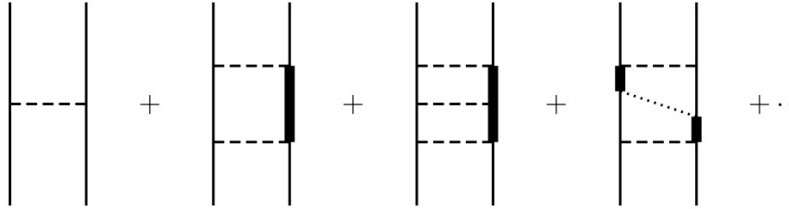}} 
\caption{\footnotesize  $\Delta$-corrected 2N force consistent with the $\Delta$-mediated 3N and 4N forces. By its iteration the $\pi \rm N \Delta$ vertex yields the $\pi$-exchange part of the $\rm N \Delta$ to $\Delta \rm N$ two-baryon potential in a time-delayed form, as in the last shown contribution; that contribution does not arise in a truncated theory without the one-baryon piece in the hamiltonian of Fig.~4; such a truncated theory is realized by the coupled-channel two-baryon potential CD Bonn + $\Delta$.}
\end{figure}

With respect to a realistic application in few-N systems, the hamiltonian of Fig.~4 was simplified.
Without active $\pi$s, i.e., without the Hilbert sector with a $\pi$ of Fig.~3 and without the one-baryon piece of Fig.~4, the hamiltonian got  truncated, but was in this truncated form tuned \cite{deltuva01} with high precision to 2N data below the $\pi$-production threshold. The resulting hamiltonian is not realistic above the $\pi$-production threshold, its $\Delta$ part remains underdetermined, but it is as realistic with respect to the account of the elastic 2N data as other high-precision 2N potentials and therefore well suited for the description of nuclear phenomena at low energies. The resulting coupled-channel two-baryon potential will be referred to as $\rm CD \, \rm Bonn + \Delta$; its purely nucleonic reference potential \cite{machleidt03} is $\rm CD \, \rm Bonn$, whose extension it is. As the full hamiltonian of Fig.~4, also that truncated hamiltonian provides consistent 2N, 3N and 4N, in general many-N forces, for what Fig.~6 shows examples; their forms and strengths are fixed, they do not allow any further tuning to 3N and 4N data, as is possible and necessary for the 3N forces of the purely nucleonic description in Subsection 3.1. Among the arising 3N-force contributions, the Fujita-Miyazawa process \cite{fujita} is the one of lowest potential order  and of greatest dynamic importance, the $\Delta$-ring process the most pronounced one of higher potential order, as also observed  \cite{illinois,krebs01} in other approaches. If the full hamiltonian of Fig.~4 were tuned also to the data of all single-$\pi$ channels, the resulting many-N forces would receive an even better theoretical support, a tuning process, equivalent to the 3N-force tuning of the purely nucleonic forces. Furthermore, the $\Delta$-mediated many-N forces of Fig.~6 are still physicswise incomplete, since other mechanisms leading to irreducible and tunable many-N forces besides the $\Delta$-mechanism are left out. 

The coupled-channel approach to low and intermediate-energy dynamics, developed in this subsection, provided full conceptual consistency between {\em genuine} 2N and 3N forces for the first time in a microscopic and realistic nuclear theory; such a consistent description is now provided by $\chi$EFT  and has become standard.  The {\em genuine} forces of the coupled-channel approach are derived from $\pi$, sigma, rho and omega meson exchanges and are therefore conceptually determined also at intermediate and short distances. 

\begin{figure}[th]
$\Large \rm \hspace{6mm} Fujita-Miyazawa \hspace{37mm} higher \,\,order \,\, 3N \,\, force \vspace{5mm}$ \\ 
\centerline{\includegraphics[width=0.9\textwidth]{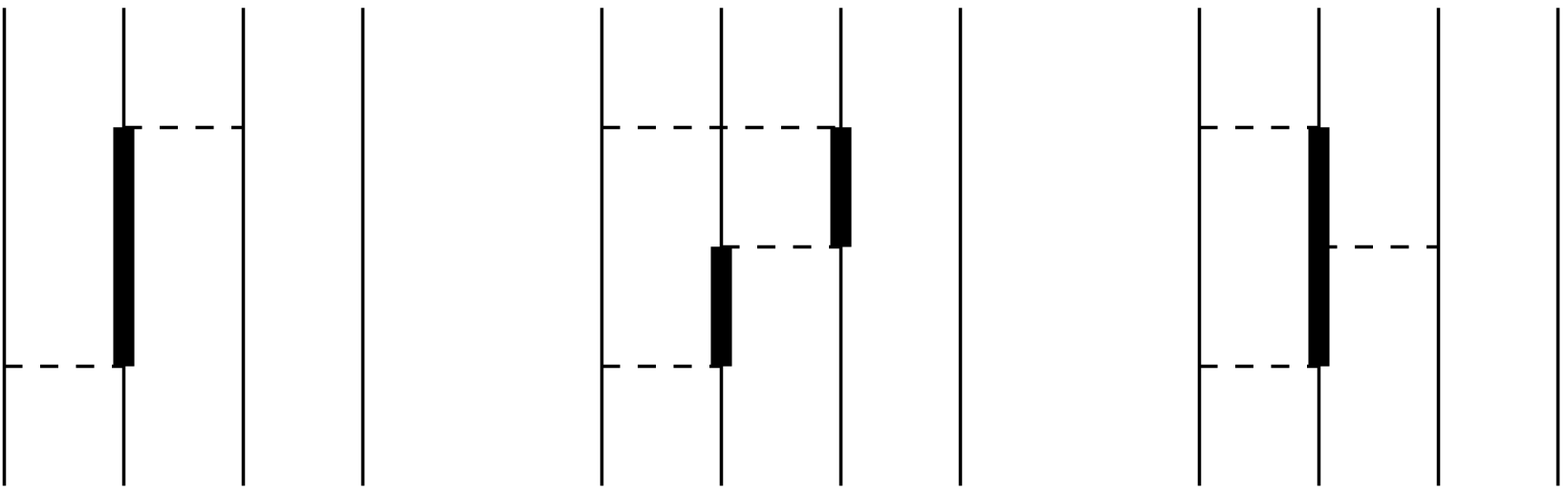}} \\  \\ \vspace{0mm}
$\Large \rm \hspace{58mm} 4N \,\, force \vspace{5mm} $ \\
\centerline{\includegraphics[width=0.6\textwidth]{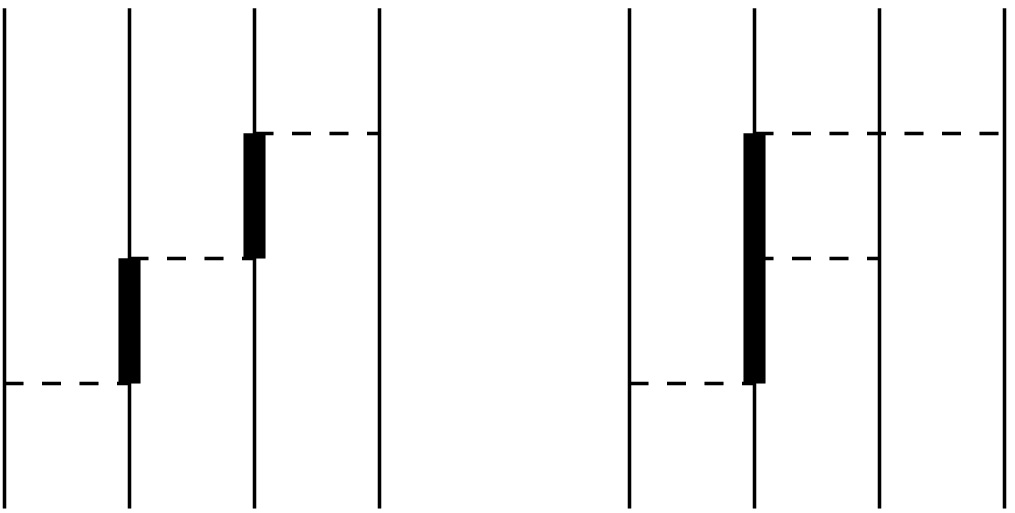}}
\caption{\footnotesize $\Delta$-mediated 3N and 4N forces, consistent with each other and with the 2N interaction. They are illustrated in a Hilbert space of baryon number 4. The upper row shows examples for the arising 3N force, the Fujita-Miyazawa process  \cite{fujita} being the one of lowest potential order; among the higher-order processes the first is the $\Delta$-ring one. The lower row shows examples for the arising 4N force. In contrast to other approaches, all possible meson exchanges are considered in the two-baryon potentials of the coupled-channel approach and therefore contribute to the many-N forces.}
\end{figure}

%%%%%%%%%%%%%%%%%%%%%%%%%%%%%%%%%%%%%%%%%%%%%%%%%%%%%%%%%%%%

\subsection{The various roles of the $\Delta$ isobar in nuclear theory} 

Due to the rather unrefined choice of its potential parameters, the coupled-channel approach of Subsection 3.2 appears old-fashioned and outdated, when viewed with modern $\chi$EFT eyes. However, the explicit inclusion of the $\Delta$ isobar into the nuclear dynamics is receiving new attention, also in $\chi$EFT; adding the $\Delta$ as a further baryon degree of freedom to a $\Delta$-full $\chi$EFT theory has practical advantages: Dynamic processes, relegated otherwise into vertices and into the low-energy constants of the $\Delta$-less $\chi$EFT, become explicit; the chiral expansion appears to converge faster due to the explicit $\Delta$ isobar; contributions to the 2$\pi$-exchange attraction and to the 3N force arise in lower chiral order than in the $\Delta$-less theory. The $\Delta$-full version \cite{krebs02,krebs02a} of $\chi$EFT sees important contributions to the 2N and 3N forces arising from the explicit account of the $\Delta$ isobar; the Fujita-Miyazawa and the $\Delta$-ring processes, illustrated in Fig.~6, turn out to be of great importance also in the $\chi$EFT approach. However, the inclusion of the $\Delta$ isobar in the present $\chi$EFT version is entirely different from the objective of the coupled-channel approach in Subsection 3.2. In $\chi$EFT the $\Delta$ isobar is considered explicitly in the underlying field theory, but the potentials are derived for use in a purely nucleonic Hilbert space and remain instantaneous. This strategy is conceptually consistent, since at present $\chi$EFT still aims at the description of nuclear properties below the $\pi$-production threshold. In contrast, the coupled-channel approach wants to go beyond the $\pi$-production threshold explicitly and therefore keeps the $\Delta$ isobar as active degree of freedom also in the Hilbert space.

However, the practical objectives of the coupled-channel potential approach and of the $\Delta$-full $\chi$EFT do not have to remain distinct for good. For example: The transition potential NN to N$\Delta$ was derived in Ref.~\refcite{deltuva01} by meson theory, but it could now be obtained in $\chi$EFT and would receive a $\pi$-exchange contribution already in leading chiral order; $\chi$EFT would also provide, besides the $\pi \rm N \Delta$ vertex, novel $\pi$-production mechanisms of baryon numbers 2 and 3, demonstrating the intimate relationship between {\em genuine} 3N and many-N forces and the mechanisms of $\pi$ production and absorption; $\chi$EFT would yield additional irreducible 3N forces besides and consistent with the $\Delta$-mediated ones; $\chi$EFT should still leave the N, in contrast to the $\Delta$ isobar, a dressed physical baryon, thereby preserving unitarity on the level of the reactions without and with a single $\pi$. The present applicability limitations of the $\Delta$-full $\chi$EFT can be overcome; $\chi$EFT could put the full coupled-channel hamiltonian of Subsection 3.2 on an improved dynamic basis and make it more attractive for wider-spread applications. Of course, that extension to intermediate energies will possibly be faced with additional convergence problems of the chiral expansion, but, on the other hand, such an enterprise could offer tremendous conceptual satisfaction and novel insight into the properties of nuclear forces and into their impact on nuclear systems.

%%%%%%%%%%%%%%%%%%%%%%%%%%%%%%%%%%%%%%%%%%%%%%%%%%%%%%%%%%%%

\section{The {\em Genuine} 3N Force in Calculations} 

I return to the two basic problems of nuclear theory, the technical problem of solving the many-body problem for the nuclear phenomena under study and the question of the conceptual validity of the chosen {\em genuine} forces, the subject of this review. Both problems are intertwined.

The conceptual question can not be settled by theoretical considerations, but is usually approached by tests of theoretical predictions, derived from chosen forces, against experimental data. Obviously, this strategy requires nuclear systems whose many-body problem is solvable with high precision, and it requires a rich amount of experimental data for the nuclear systems studied. Nuclear systems qualifying for those tests are 3N and 4N bound and scattering states, the bound states and narrow resonances of light nuclei heavier than the 3N and 4N ones,  and nuclear matter, the artificial system with an infinite number of Ns reflecting binding and saturation properties of heavy nuclei. The starting point are always calculations with the simplest form of realistic {\em genuine} forces, those of 2N nature; if those calculations are unable to account for the existing data with satisfying accuracy in the beginning round, a change of the chosen 2N potential to another 2N one is often attempted; if that also fails, one resorts to the addition of {\em genuine} forces of next-higher complexity, i.e., first of a 3N force. \\

3N and 4N systems are the best studied nuclear systems; they are considered the gold standards for testing assumed forces between Ns. Their many-body problem is for bound and scattering states conceptually under control due to Faddeev \cite{faddeev} and Alt, Grassberger and Sandhas \cite{alt} (AGS),  and it is getting, step-by-step, also calculationally under control by various numerical techniques. Each of the theory groups, embarked in calculations of 3N and 4N systems, chooses its particular calculational scheme and, as a consequence, has its favorite form of dynamics. I give examples:

\begin{itemize}
\item
My collaborators and me adopted AGS integral equations in momentum space as our numerical technique. The calculations are quite tricky for few-N scattering due to singularities, though the singularities  are integrable; they arise from open inelastic channels. The latest important technical achievements were the inclusion of the Coulomb interaction between protons (p) in the momentum-space scattering equations \cite{deltuva02}  and the calculation of 4N scattering \cite{deltuva03a,deltuva03b} above the three- and four-body breakup thresholds. The inclusion of Coulomb was a stumbling block for the theoretical description in momentum space during decades; Coulomb is important at low energies, at all energies in forward direction  and in certain kinematics of breakup reactions; its proper treatment is necessary to uncover the role of nuclear forces in the  studied observables. We use the technique of screening and renormalization; it fails at thresholds for charged two-body clusters where Coulomb overwhelms all other interactions;  it is also not applicable for the breakup into three observed charged bodies in $\rm p{^3}He$ reactions. Our technique uses the purely nucleonic potential CD Bonn and its coupled-channel extension CD Bonn + $\Delta$  as dynamics, the latter generating consistent 2N, 3N and 4N forces as described in Subsection 3.2; the technique is not yet extended to also cope with general irreducible 3N forces besides the ones, mediated by the $\Delta$ isobar.
\item
The approach of Ref.~\refcite{krakau01} uses integral equations of the Faddeev type in momentum space; it can deal routinely \cite{krakau02} with an irreducible 3N force; at present, it  seems to work preferably with  $\chi$EFT potentials; it can deal with breakup reactions; Coulomb is not incorporated yet; it is also not yet extended to describe 4N reactions. Ref.~\refcite{pisa01} uses the Kohn variational principle in connection with the hyperspherical-harmonics expansion of the coordinate-space scattering wave functions, whose asymptotic form has to be imposed. The technique includes Coulomb, is applicable to local and nonlocal 2N and 3N nuclear potentials, though the local version \cite{navratil01} of the $N{^2}LO$ 3N force comes as a handy simplification, and it can be employed for 3N and 4N reactions with two bodies in the final state; it can go beyond breakup thresholds, but has not yet been extended to the breakup reactions themselves. There are other groups with further alternative 3N and 4N techniques, e.g., the ones of Refs.~\refcite{japan} and \refcite{grenoble}, with their individual advantages, but also their special limitations. 
\end{itemize}

\noindent
In a collective effort of different techniques, the few-body community is now able to cover the 3N and 4N systems computationally in all their low-energy aspects. Benchmark checks between the groups guarantee that all numerical techniques employed are reliable in their realm of applicability. However, all results shown later on in this review are exclusively obtained by our technique, dynamically based on the coupled-channel potential CD Bonn + $\Delta$, which I am best familiar with; the technique has also the widest range of applicability. I shall always point out to what extent the shown results are general or indicate important deviations with respect to the results of other groups. In contrast, Refs.~\refcite{krakau02} and \refcite{pisa01} can present results for traditional high-precision 2N and 3N potentials and for whole families of 2N and 3N  $\chi$EFT potentials. On the experimental side, there is a multitude of data, especially now data of reactions with polarized particles. From those data one can hope to get information on nuclear forces, piece by piece. I describe important steps of that project in this review. \\

I have been a great fan of few-nucleon physics and a convinced believer that  3N and 4N systems exclusively provide the key for learning about the detailed properties of nuclear forces. I have to acknowledge that reliable {\em ab initio} calculations of finite nuclei, heavier than the 3N and 4N systems, have become possible, and they are accurate enough to  provide complementary information on nuclear forces. Light nuclei, their bound states and their narrow resonances, treated as being bound, are other systems, able to effectively test the nuclear forces; the study of reactions in those nuclei is under way.  Quantum Monte Carlo \cite{wiringa}, bound to the use of local potentials, and no-core shell-model \cite{nocore01} calculations, combined with SRG-evolved $\chi$EFT potentials, are successful techniques for solving the corresponding nuclear many-body problems; the coupled-cluster approach \cite{coupledcluster} can carry microscopic calculations to even heavier nuclei.  In the early days of microscopic nuclear-structure calculations, nuclear matter was the favorite testing ground for realistic {\em genuine} forces; however, an accurate treatment of the arising 3N-cluster contributions in the framework of Brueckner theory created enormous technical problems which present-day calculations \cite{heberler} are able to overcome in part. 

%%%%%%%%%%%%%%%%%%%%%%%%%%%%%%%%%%%%%%%%%%%%%%%%%%%%%%%%%%%%

\subsection{ 3N and 4N binding energies} 

2N forces underbind 3N and 4N bound states, as shown in Table~1; that is a general result, also obtained by all other groups. The miss is emphasized by its comparison with the experimental binding energy, which is the sum of two large and quite force-dependent  contributions of opposite sign, of the potential energy, e.g., for the 3N system being of the order of - 50 MeV, and of the kinetic energy, e.g., for the 3N system being of the order of + 45 MeV; on this scale, the 3N binding contribution by the 3N force of about 1 MeV is small, just a 2\% effect. When employing the purely nucleonic potential CD Bonn and its coupled-channel extension CD  Bonn + $\Delta$  as in Table~1, an additional tuning to 3N and 4N data is impossible, quite in contrast to other approaches with many-N forces; thus, our technique carries the binding-energy miss over to the theoretical description of scattering. In other approaches, the binding of the 3N bound states is used as a data point for tuning the 3N force; the 4N binding correlates \cite{tjon} with the 3N binding and is therefore not a fully independent observable. The binding energy difference of the mirror nuclei $\rm {^3}H$ and $\rm {^3}He$, due to Coulomb and the charge asymmetry of nuclear forces, are theoretically well accounted for.

\begin{table}[pt]
\tbl{Binding energies of 3N and 4N bound states.}
{\begin{tabular}{@{}lccc@{}} \toprule
\hphantom{Piston mass} & ${}^3\rm H$ &  ${}^3\rm He$ &  ${}^4\rm He$ \\
& (MeV) & (MeV) & (MeV) \\ \colrule
CD Bonn & \hphantom{-}8.00 & \hphantom{-}7.26 & 26.18 \\
CD Bonn + $\Delta$ & \hphantom{-}8.28 & \hphantom{-}7.54 & 27.10 \\
exp & \hphantom{-}8.48 & \hphantom{-}7.72 & 28.30 \\ \colrule
$\Delta \rm E_2$ & -0.51 & -0.48  & -2.80  \\
$\Delta \rm E_{3}$(FM) & \phantom{-}0.50 & \hphantom{-}0.48 & \hphantom{-}2.25 \\
$\Delta \rm E_{3}$(h.o.) & \hphantom{-}0.29 & \hphantom{-}0.28 & \hphantom{-}1.30 \\
$\Delta \rm E_{4}$ & & & \hphantom{-}0.17 \\
\botrule
\end{tabular}}
\begin{tabnote}
The results are taken from Ref.~\refcite{deltuva04}. Compared with the purely nucleonic reference potential CD Bonn, the $\Delta$-mediated effects are of 2N, 3N and 4N nature, denoted by $\Delta \rm E_2$, the 2N dispersive effect, and by $\Delta \rm E_{3}$ and $\Delta \rm E_{4}$, the 3N- and 4N-force effects. The 3N-force contributions $\Delta \rm E_{3}$ are further split into contributions arising from the Fujita-Miyazawa (FM) and from higher-order (h.o.) mechanisms, as illustrated in Fig.~6.
\end{tabnote}
\end{table}

The 3N- and 4N-force contributions to binding are in general small, the 4N-force contribution even an order of magnitude smaller than the already small 3N-force contribution. The 4N-force contribution, shown in Table~1, is in qualitative agreement with the finding of Ref.~\refcite{nogga}, which is obtained, in perturbation theory, with the 4N force of Ref.~\refcite{epelbaum03c}. The long-standing folklore about the hierarchy of the importance of many-N forces, i.e., the decreasing importance with increasing number of involved Ns, was put on a solid theoretical basis by $\chi$EFT and its power counting, and it is confirmed by the theoretical results  of 3N and 4N binding energies.

%%%%%%%%%%%%%%%%%%%%%%%%%%%%%%%%%%%%%%%%%%%%%%%%%%%%%%%%%%%%

\subsection{3N and 4N scattering} 

There is a large amount of {\em 3N scattering} data, especially also data for reactions with polarization. Most data, by far the majority, can be described well by a dynamics based on 2N forces alone; the reason is that the 3N system is a rather low-density system in scattering. Furthermore, the 2N force is exploited by the 3N low-energy observables not far from its elastic on-shell limit; the spread of results arising from different realistic 2N potentials is minor. \\

3N-force effects on low-energy observables are small. There remain only  few low-energy 3N puzzles, i.e., discrepancies between data and theoretical predictions, unresolved at present. Of course, the puzzles are the conceptually even more interesting situations compared with the multitude of successful descriptions of data. The puzzles, most looked at, are the N analyzing power  $\rm A_y$  of elastic Nd scattering  \cite{deltuva05} below 20 MeV N lab energy, not illustrated by a plot in this review, and the pd and neutron-d (nd) breakup reactions in  the space-star kinematics at 13 MeV N lab energy, illustrated in Fig.~7; both puzzles appear insensitive to the presently available 3N forces. 

\begin{itemize}
\item The few-nucleon community got surprised by the $\rm A_y$ puzzle, which has been remeasured several times; the data are stable and appear therefore beyond any consistency doubt. Theoretically, the spin observable $\rm A_y$ is complicated due to competing contributions.  The recent calculational effort \cite{witala01,witala02} was done even with the  inclusion of the most advanced  $\chi$EFT 3N force, almost complete up to  the chiral order $N{^3}LO$, and was nevertheless unable to resolve this puzzle.
\item The few-body community expected a sizable 3N-force effect in breakup for the space-star kinematics on intuitive grounds; in this kinematics, the three Ns leave the interaction region in a star formation and therefore appear, classically, to exploit the 3N force most efficiently; this is the historic reason why the experiments were originally done, and this is why data and the theoretical predictions are shown in the present context. However, for those breakup data, there is, after the first round of calculations, no indication \cite{deltuva05, witala02,japan} for such a 3N-force signal; furthermore, the data for nd and pd breakup are far apart; if upheld, they indicate a strong nuclear charge-asymmetry effect, since the Coulomb effect  \cite{deltuva05,japan} appears small, though in other Nd reactions there is no indication for such a strong nuclear charge asymmetry. The nd breakup was remeasured several times, thus, they appear validated well; the pd data got also confirmed by one additional measurement. 
\end{itemize}

\begin{figure}[th]
\centerline{\includegraphics[width=0.85\textwidth]{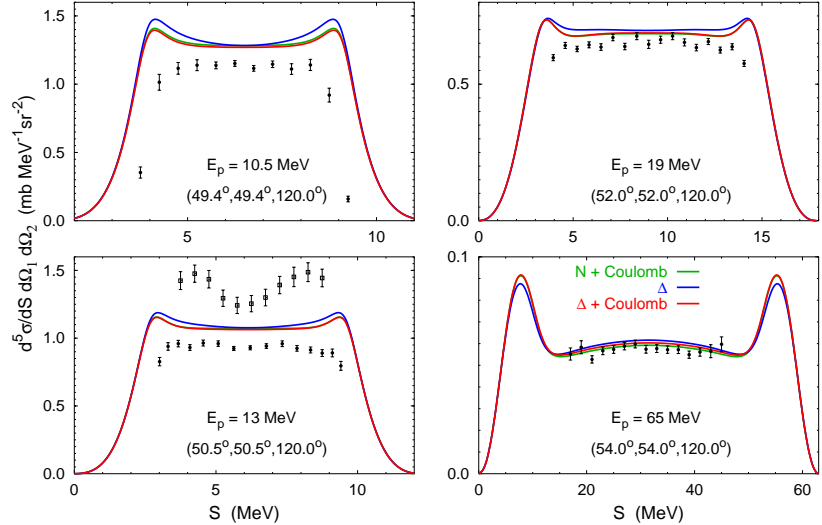}} 
\caption{\footnotesize Nd break-up cross sections in the the space-star kinematics for N lab energies from 10 MeV to 65 MeV along the kinematical locus S. The green (red) line label N ($\Delta$) + Coulomb indicates the results obtained from the CD Bonn (CD Bonn + $\Delta$) potential with the inclusion of the Coulomb interaction between the ps; the difference between the two lines demonstrates the 3N-force effect on the observables, practically invisible in the plot. The blue line $\Delta$ gives the results without Coulomb, representing our nd prediction; its difference to the red line $\Delta$ + Coulomb  illustrates the Coulomb effect, which is also not important for these observables. The theoretical results are taken from Ref.\cite{deltuva05} , in which also the references to the experimental data are given, except for one additional confirmation \cite{sagara} of  the pd data at 13 MeV.}
\end{figure} 

In contrast to low-energy scattering observables, sizable 3N-force effects are seen at higher energies, i.e., in the diffraction minima and in some spin observables of Nd elastic scattering. Fig.~8 shows examples, the diffraction minimum of pd elastic scattering at 135 MeV p lab energy and some spin observables of elastic pd scattering at that energy and at higher energies. It is gratifying that the 3N force provides a strong effect in the diffraction minimum going into the right direction, as in other calculations. The 3N force has always to compete with 3N correlations, built from successive 2N-force contributions. In my view, 3N-force effects do not show up at higher energies in general, they show up, whenever the 2N-force contributions in 3N correlations are accidentally small as happening in the observables and at the particular energies of Fig.~8. There is also a beneficial 3N-force effect on the spin observables of Fig.~8; it is still too early to judge these results a convincing success of the employed 3N force or just a lucky accident. \\

\begin{figure}[th]
\vspace{0mm}
\centerline{\includegraphics[width=0.40\textwidth]{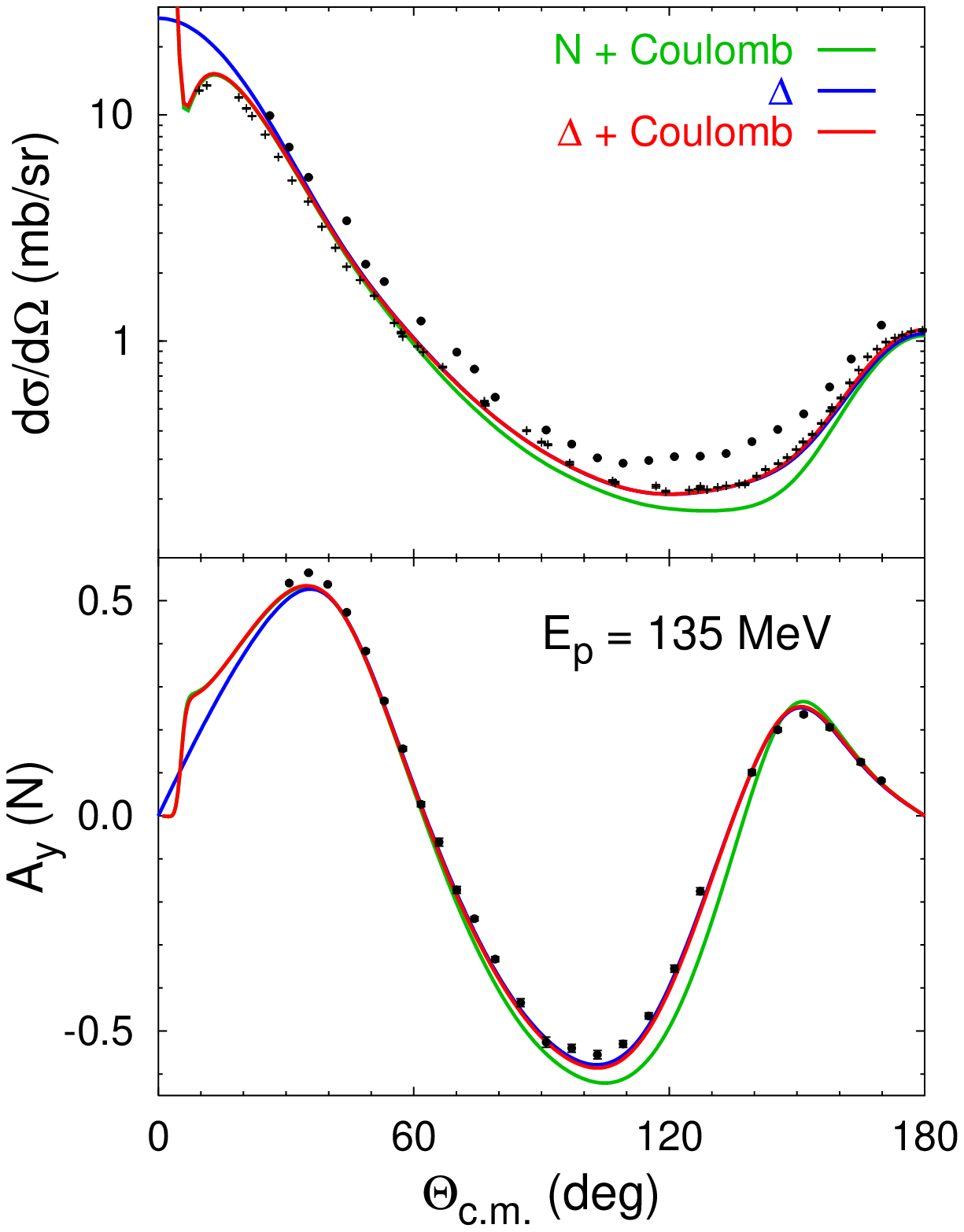} 
                    \includegraphics[width=0.40\textwidth]{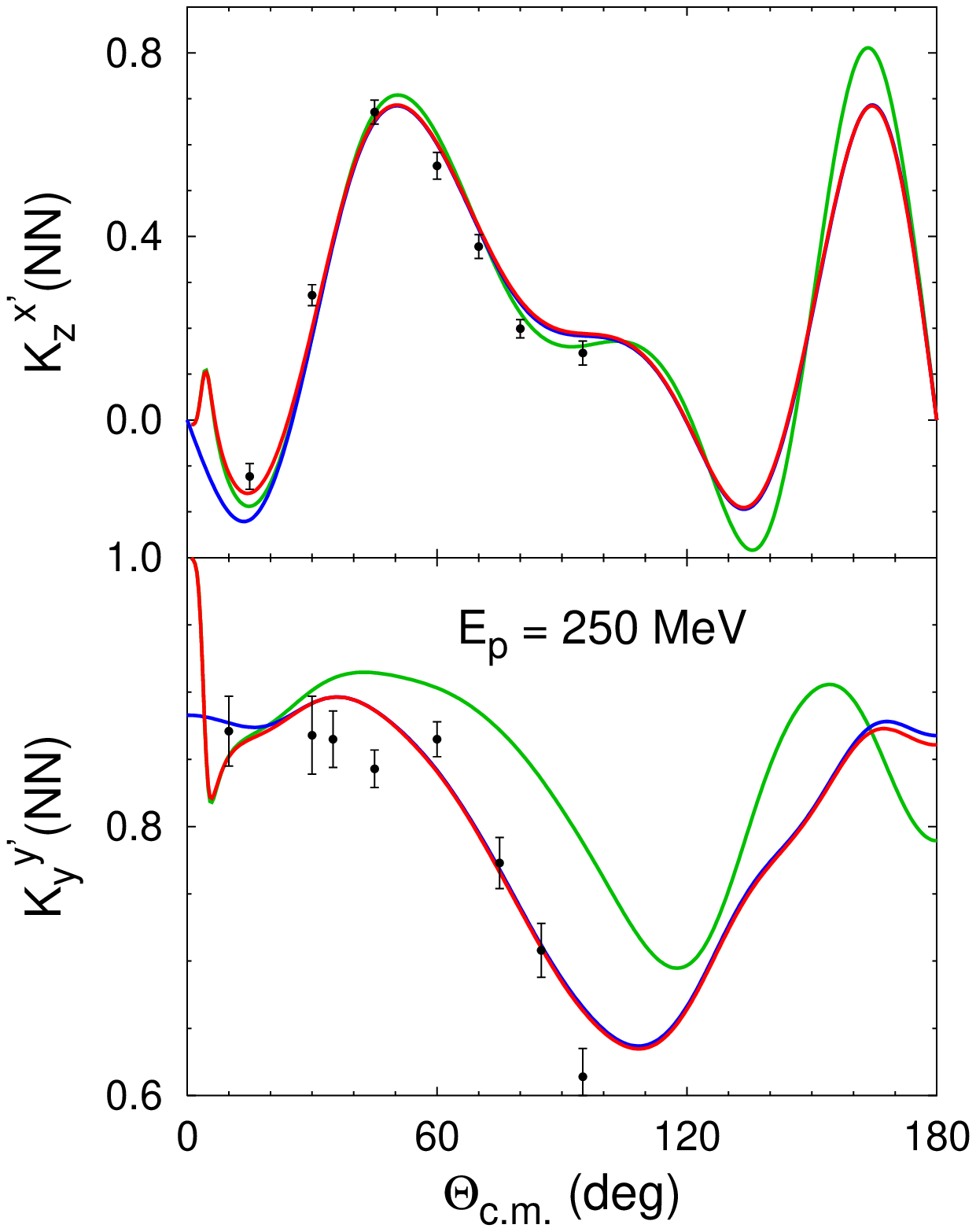}}
\caption{\footnotesize Selected observables of pd elastic scattering at higher energies as function of the c.m. angle $\Theta_{\rm c.m.}$. The green (red) line label N ($\Delta$) + Coulomb indicates the results obtained from the CD Bonn (CD Bonn + $\Delta$) potential with the inclusion of the Coulomb interaction between the ps; the difference between the two lines demonstrates the 3N-force effect on the observables. The blue line $\Delta$ gives the results without Coulomb, its difference to the red line $\Delta$ + Coulomb  illustrates the Coulomb effect. The effect of the 3N force is quite pronounced. In contrast, an effect of the Coulomb repulsion between the protons is only seen in the extreme forward direction. At 135 MeV p lab energy, the upper experimental data are from Ref.\cite{ermisch} , the lower data in the differential cross section from Ref.\cite{sekiguchi} , an example for conflicting experimental data; the theoretical predictions are from Ref.\cite{deltuva06} . At 250 MeV p lab energy,  the theoretical predictions are from Ref.\cite{deltuva07} , in which also the references to those experimental data are given.}
\end{figure} 

{\em 4N reactions} are not as widely explored as 3N reactions. Nevertheless, the 4N system is also in scattering denser, rich of low-energy resonances and therefore more likely to  show 3N-force effects. An important example is the total $\rm n {^3}H$ cross section in the resonance region around 3.5 MeV c.m. energy. Whereas in  3N scattering the particular form chosen for 2N and 3N forces does not really matter strongly for theoretical predictions, in 4N scattering it does.  Our result for the total $\rm n {^3}H$ cross section, obtained from the CD Bonn and CD Bonn + $\Delta$ potentials, is shown in Fig.~9; the 3N-force effect is quite sizable, but goes into the unwanted direction of decreasing the resonance peak; this has been a general result for traditional forces. In contrast, this trend is broken by the predictions derived in Ref.~\refcite{pisa02} with $\chi$EFT potentials of chiral order $N{^3}LO$ for the 2N force, but only of chiral order $N{^2}LO$ for the 3N force, used in the local version of Ref.~\refcite{navratil01}; according to Ref.~\refcite{pisa03}, the latter $\chi$EFT 3N force is much softer than traditional 3N forces; the different $\chi$EFT result for the total $\rm n {^3}H$ cross section in the resonance region around 3.5 MeV c.m. energy may be traced back to that property.

\begin{figure}[th]
\vspace{0mm}
\centerline{\includegraphics[width=0.60\textwidth]{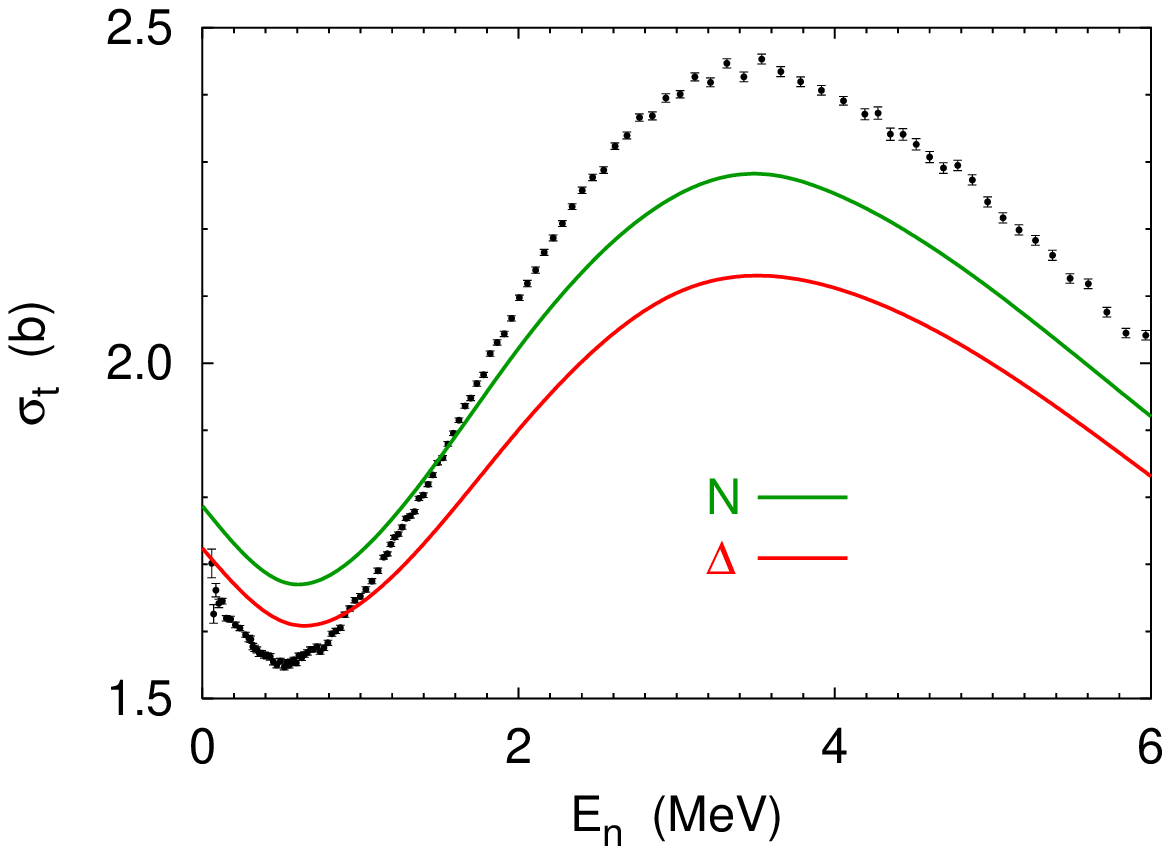}} 
\caption{\footnotesize Total $\rm n{^3}H$ cross section as function of the n lab energy from threshold to 6 MeV. The green (red) line label N ($\Delta$) refers to the results obtained from the CD Bonn (CD Bonn + $\Delta$) potential; the difference between the two lines N  and $\Delta$ demonstrates the 3N-force effect on the observable. The theoretical results are from Ref.\cite{deltuva04} , in which also the references to the experimental data are given.}
\end{figure} 

\begin{figure}[th]
\vspace{0mm}
\centerline{\includegraphics[width=0.50\textwidth]{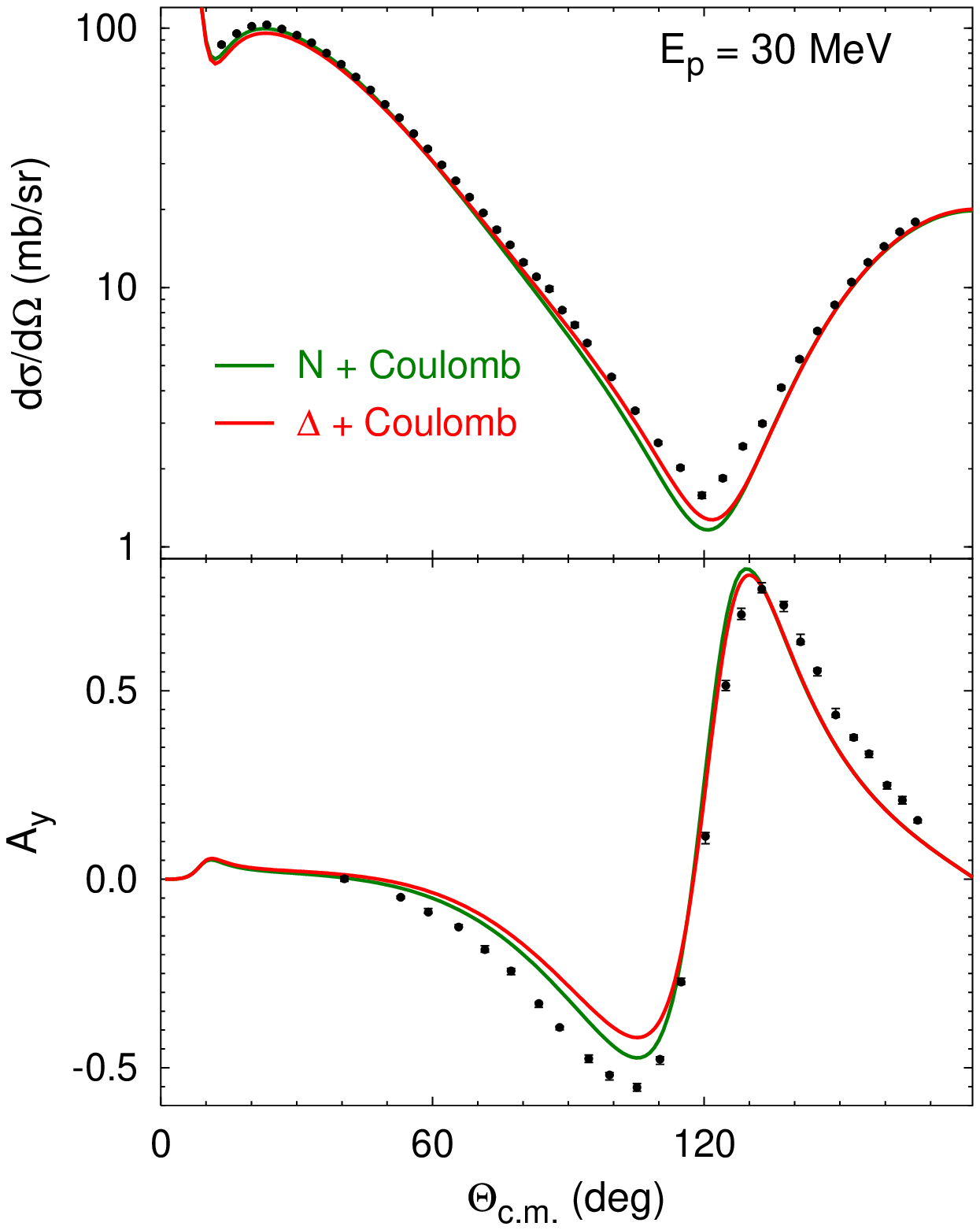}} 
\caption{\footnotesize Elastic $\rm p{^3}He$ differential cross section and p analyzing power $\rm A_y$ at 30 MeV p lab energy as function of the c.m. angle $\Theta_{\rm c.m.}$. The green (red) line label N ($\Delta$) + Coulomb indicates the results obtained from the CD Bonn (CD Bonn + $\Delta$) potential with the inclusion of the Coulomb interaction between the ps; the difference between the two lines demonstrates the 3N-force effect on the observables. The theoretical results are from Ref.\cite{deltuva03b} , in which also the references to the experimental data are given.}
\end{figure} 

Another important low-energy 4N scattering observable is elastic $\rm p{^3}H$ scattering, below the $\rm n{^3}He$ threshold; it is the energy region in which the first excited $\rm {^4}He$ state is embedded in the $\rm p{^3}H$ continuum, close to threshold. The theoretical prediction of Ref.~\refcite{lazauskas} observes a strong dependence of the resonance properties on the proper description of the thresholds, and it observes important 3N-force effects; due to the applicability limitation of our Coulomb treatment, this observable is out of our reach at present. Ref.~\refcite{leidemann} sees the same resonance in inelastic electron scattering from  $\rm {^4}He$ and also a dramatic sensitivity of the theoretical transition form factors on the 2N and 3N forces. 

The p analyzing powers $\rm A_y$ of p${^3}\rm He$ elastic scattering at 10 MeV p lab energy \cite{deltuva04,deltuva03b} and at energies below \cite{kievsky02} , is as difficult to describe as the corresponding analyzing power of Nd elastic scattering; those 3N and 4N observables are both the results of competing contributions, though the individual contributions are not identical in the 3N and 4N systems. As for the total $\rm n {^3}H$ cross section in the resonance region around 3.5 MeV c.m. energy, Ref.~\refcite{kievsky02} is able to account for the 4N $\rm A_y$ with $\chi$EFT potentials of chiral order $N{^3}LO$ for the 2N force, but only of order $N{^2}LO$ for the 3N force, used in its local version of Ref.~\refcite{navratil01}; unfortunately, the same force combination cannot resolve the 3N $\rm A_y$ puzzle. Most other spin observables in elastic 4N scattering at low energies, observed till now, are rather uneffected by a 3N force and are described quite well by 2N forces alone. 

Above the three- and four-body breakup thresholds, there is a 3N-force effect in the cross-section minima of elastic 4N scattering. Fig.~10 shows a small effect in the diffraction minimum of $\rm p ^{3}\rm He$ elastic scattering at 30 MeV p lab energy; in fact, the isolated 3N-force effect is larger than the total one, since it is balanced by the 2N dispersion. This fact is also observed in 3N scattering at most energies, but the 3N-force effect becomes quite visible there at higher energies, as demonstrated in Fig.~8. At higher 4N scattering energies, one can therefore hope to also find a region, as in 3N scattering, in which the 2N-force contributions are small and therefore do not cover up 3N-force effects. Fig.~10 also shows that at higher energies the p analyzing power $\rm A_y$ is described well in its peak, but that it is overestimated in its minimum, a 3N-force effect going into the unwanted direction. \\

The 3N-force contribution is required for the fine-tuning of binding energies, therefore in many-N reactions for the proper position of thresholds, and it is seen in resonances. Otherwise, the existing 3N and 4N data see the signal of a 3N force only in rather exceptional cases; the effect can be stronger in 4N than in 3N reactions. Refs.~\refcite{deltuva03b} and \refcite{deltuva04} also separate 3N- and 4N-force contributions to the observables of Figs.~9 and 10 for 4N scattering, as was done in Table~1 for 3N and 4N binding. The dominant 3N-force contribution is from the Fujita-Miyazawa process; the $\Delta$-ring process is the largest one of those labeled higher order; the 3N-force contribution has to balance the rather sizable 2N dispersion, arising for the $\Delta$-mediated nuclear forces. The 4N-force effect is minute, even when compared with the small 3N-force contributions.

%%%%%%%%%%%%%%%%%%%%%%%%%%%%%%%%%%%%%%%%%%%%%%%%%%%%%%%%%%%%

\subsection{Light nuclei: ground states, narrow resonances and scattering} 

The bound states and narrow resonances of p-shell nuclei are systematically studied in quantum Monte Carlo calculations \cite{wiringa,pieper} with traditional local high-precision 2N  potentials, often combined with the Illinois \cite{illinois} 3N force, and in the no-core shell model \cite{nocore01} with $\chi$EFT-based 2N and 3N forces, additionally softened by the SRG evolution. The theoretical  account of the experimental level schemes is impressive. The need of the inclusion of a 3N force is obvious, especially for the proper level ordering. E.g., the two different types of calculations  \cite{pieper,navratil02} achieve both the proper level ordering in the nucleus $^{10}\rm B$, and they do so with different 3N forces. Furthermore, both observe the importance of  $\Delta$-like contributions to the 3N force, especially the Fujta-Miyazawa and $\Delta$-ring processes. In the present context, the $\Delta$-like contributions are employed in instantaneous and local form in distinction to the coupled-channel approach of Ref.~\refcite{deltuva01} with explicit $\Delta$s. In contrast to the 3N and 4N results, the $\Delta$-ring process is important in p-shell nuclei due to its isospin-3/2 contribution in 3N configurations. 

The additional SRG softening of the employed 2N and 3N $\chi$EFT  forces makes microscopic and realistic calculations for nuclei, even heavier than $\rm ^{16}O$,  tractable. The no-core shell-model treatment then converges sufficiently rapidly, despite the rapidly growing number of contributing nuclear configurations. A particularly nice example for the achieved success, is the proper account  \cite{otsuka,hergert} of the binding of the $\rm O$ isotopes: The 3N force yields effective repulsive contributions to the interactions among the excess ns with the Ns of the $^{16}\rm O$ core and thereby leads to the experimental n drip line at $^{24}\rm O$.

The microscopic description of nuclear reactions has been considered the exclusive realm of 3N and 4N scattering theories. This situation is changing. Ref.~\refcite{wiringa02} demonstrates how quantum Monte Carlo with local 2N and 3N potentials and Ref.~\refcite{navratil03}  how the no-core shell model with 2N and 3N $\chi$EFT potentials, SRG-evolved, can be extended to scattering calculations in light nuclei; the no-core shell model is combined  with the resonating group technique for scattering. Elastic N${^4}\rm He$ scattering up to the d${^3}\rm H$ threshold is the first instructive example, showing important 3N-force effects on the $J{^\pi}$ 3/2$^-$ and 1/2$^-$ resonances, on their spin-orbit splittings,  and on differential cross sections and the N analyzing power $\rm A_y$; this finding is similar to the strong 3N-force effect on the low-energy resonance in elastic n${^3}\rm H$ as discussed in the context of  Fig.~9. The overall agreement, achieved with the help of {\em genuine} 3N forces, is impressive. Ref.~\refcite{wiringa02} achieves a better agreement for the 3/2$^-$ resonance; though the qualitative findings are beyond doubt, the full calculational convergence of the scattering results of the no-core shell model is not yet secured. A benchmark comparison of elastic n${^3}\rm H$ with the corresponding 4N scattering result, calculationally confirmed,  were desirable.

%%%%%%%%%%%%%%%%%%%%%%%%%%%%%%%%%%%%%%%%%%%%%%%%%%%%%%%%%%%%

\subsection{Nuclear matter}

In the early days of microscopic nuclear structure with realistic 2N forces,  on which I reflected in Section 1, nuclear matter, with its properties energy per nucleon and saturation density, was the system of choice for a quantitative check of the forces employed in nuclear theory. The practical stumbling block of the employed Brueckner theory was the calculation of the 3N cluster \cite{bethe01,bethe02} which, in Section 1, I declared the first {\em induced} 3N force encountered in nuclear theory. The early computational results were polluted by so large theoretical uncertainties that the nuclear-matter problem was deserted by some nuclear theorists, including myself; instead, few-N systems appeared less complicated and therefore more fruitful for studying nuclear forces. Even after so many years, the calculation of symmetric nuclear and neutron matter are still marred by sizable computational errors. Nevertheless, two important results on the role of {\em genuine} nuclear forces are emerging:

\begin{itemize}
\item The inclusion of a well-founded {\em genuine} 3N force \cite{pandhi}  is most important for the prediction of the correct nuclear-matter saturation properties.
\item Realistic saturation properties \cite{heberler,kohno} can be achieved, even if the {\em genuine} 2N and 3N forces are rather soft as the $\chi$EFT potentials, presently in use, and even when the employed 3N force is fitted just to data of few-N systems, e.g., to the ${^3}\rm H$ binding and the ${^4}\rm He$ charge (point-charge) radius, and not to nuclear matter itself.
\end{itemize}

\noindent
Since nuclear-matter calculations still fight with sizable uncertainties, nuclear matter has not yet regained an important place in the theoretical studies of the {\em genuine} nuclear forces; that situation is likely to change soon.

%%%%%%%%%%%%%%%%%%%%%%%%%%%%%%%%%%%%%%%%%%%%%%%%%%%%%%%%%%%%

\section{Conclusions, Concerns and Possible Strategies} 

The review discussed the difference between {\em genuine} and {\em induced} nuclear forces, but concentrated on {\em genuine} forces with the focus on 3N forces. It emphasized that also {\em genuine} forces are not measurable observables; they are part of a chosen theory concept. When nevertheless loosely talking of a signal for a 3N force in experimental data, the failure of the lowest dynamic-order description of data with a chosen 2N force is meant; the next-order approximation, which becomes necessary and includes a 3N force, depends on the lowest-order form with its 2N force; it is therefore not unique, and it is suggested by data only within a particular theory model. \\

The strategy for testing the choice of nuclear dynamics and the possible need, in the above sense,  for the addition of a {\em genuine} 3N force proceeds in three steps:

\begin{enumerate}
\item
In the first step the active hadronic degrees of freedom are chosen, their field-theoretic interactions are specified and instantaneous hermitian potentials for a many-body hamiltonian and for its use in non-relativistic quantum mechanics are derived. At present, $\chi$EFT is considered the conceptually best founded and most advanced procedure for the construction of the potentials between the nuclear constituents at low energies below the $\pi$-production threshold.
\item
In the second step nuclear many-body systems are selected whose properties can be described theoretically with high precision. The systems under study are the 3N and 4N bound states and reactions without and with polarization, the bound states and narrow resonances of light nuclei, heavier than the 3N and 4N systems, and nuclear matter.
\item
In the third step the theoretical predictions are compared with experimental data. 
\end{enumerate}

\noindent
The {\em first} step is described in Section 3 in detail and with its various options. A future attempt is suggested to put the coupled-channel approach of Subsection 3.2 with explicit $\Delta$ isobars in the Hilbert space on the firm  conceptual basis of a $\Delta$-full $\chi$EFT. The coupled-channel approach was the first nuclear force model with consistent 2N and many-N forces; its 2N part was tuned to the 2N data up to $\pi$-production threshold with the same high precision as standard potentials; its 3N part contains the Fujita-Miyazawa and the $\Delta$-ring processes, seen as prominent contributors, even  with short-range modifications due to meson exchanges beyond the $\pi$; that was a remarkable early achievement. \\

\noindent
A large amount of computational effort went into the solution of the 3N and 4N many-body problems for the {\em second} step; and that effort was technically highly successful. However, as long as the observed 3N-force effects remain rather small in those systems, then even the technically best calculations of 3N and 4N systems are not as helpful as originally hoped in yielding information on the 3N force. Instead, systems, which are still troubled by numerical challenges, but show sizable 3N-force effects as nuclei, heavier than the 3N and 4N systems, and nuclear matter, may be more informative on them, as long as still existing numerical errors do not cover up the 3N-force effects. For example, components of the 3N force with total isospin 3/2 are not operative in the 3N system and in the 4N bound state, as long as Coulomb and the charge dependence of the nuclear forces are not considered, and they are suppressed even in 4N scattering, but show noticeably up in nuclei with n-excess. Quantum Monte Carlo \cite{wiringa} and no-core shell-model \cite{nocore01} calculations, the latter combined with the SRG evolution of the chosen {\em genuine} forces, are able to explore nuclei, heavier than the 3N and 4N systems, and are providing highly important, additional information on the need for 3N forces.\\

\noindent
On the {\em experimental side} of the testing strategy, the {\em third} step requires a rich, well-checked and confirmed data base. As seen in the differential cross section of pd elastic scattering at 135 MeV p lab energy of Fig.~8, such a confirmed data base does not exist in 3N and 4N systems beyond breakup thresholds; though remeasured several times, the two data sets of the two experimental groups remain distinct. There is a multitude of 3N scattering data, without and with polarization, but a theoretical confirmation of the experimental consistency  of those data is not established yet in a manner done in 2N scattering  with the help of phase shift analyses below the $\pi$-production threshold. Indeed, there is an early, but practically unsuccessful attempt  \cite{noyes} establishing just such an approach; there are phase-shift analyses for elastic few-N scattering below their respective breakup thresholds, but not above, e.g., for pd scattering in Ref.~\refcite{psa01}, for p$^{3}\rm He$ scattering in Ref.~\refcite{psa02} and for p$^{4}\rm He$ scattering in a phase-shift analysis quoted in Ref.~\refcite{navratil03}. In the light of the existing 3N and 4N puzzles a consistency check of data at all energies up to their $\pi$-production thresholds were extremely useful. 
\\

\noindent
On the {\em theoretical side} of the testing strategy, the numerical steps from a chosen hamiltonian to the solution of the many-N system and its observables are highly complex and therefore not transparent; that situation does not help our intuitive understanding of details in the {\em third} step and therefore of the driving physics behind the computed observables. One has to admit with some resignation, that, in these days of heavy computations, only the computers seem to understand the dynamic  processes in detail. \\
 
All practical calculations with 2N and many-N {\em genuine} forces show that indeed the 2N-force contribution is most pronounced and quite sufficient for an accurate account of many data; if insufficient, the 2N-force contribution describes  at least  the gross features of data; {\em 3N-force corrections are small}, though not at all unimportant, since they are  sometimes needed for a detailed account of data, as described in Section 4. The 3N-force contribution has always to compete with the 3N correlations arising from successive 2N interactions; only when the latter ones are small, the 3N-force contribution is augmented as happens in exceptional cases, e.g., in the diffraction minimum of pd elastic scattering at 135 MeV p lab energy. Till now, the comparison of theoretical predictions with experimental data has  not provided detailed quantitative information on the required properties of the {\em genuine} 3N forces.

With respect to 4N-force effects, Ref.~\refcite{deltuva04} reports on the first full calculations for 4N bound and scattering states with consistent 2N, 3N and 4N forces. The 4N-force contributions are even an order of magnitude smaller than the 3N-force contributions. The $\chi$EFT prediction, that the importance of many-N forces decreases with the number of Ns involved due to the chiral expansion, appears well established for the nuclear densities, encountered at low energies. 3N-force corrections are the most important many-N contributions to be taken into account, and among them the one arising from the Fujita-Miyazawa process is most prominent, the $\Delta$-ring process is the most noticeable one among the other processes. In contrast, the worry on the possible practical significance of 4N forces can safely put aside at present.

In the light of the remaining 3N and 4N puzzles, i.e., of persistent disagreement between experimental data and theoretical predictions, one wonders, if  {\em phenomenological} 3N forces exist, which could be identified directly from data and which are able to resolve the known puzzles, without destroying the broad agreement for other data. Such a project is interesting, if a unique answer resulted, even if the phenomenological 3N force may be conceptually not understood yet and  may even be inconsistent with the underlying 2N force.  But due to the complexity of step three in the strategy for testing a chosen nuclear dynamics, that seemingly naive project is quite complicated. Nevertheless, this strategy has been attempted: 

\begin{itemize}
\item In the light of the $\rm A_y$ puzzle, Ref.~\refcite{kievsky} made the interesting suggestion for a strong spin-orbit 3N-force component, a suggestion by and large ignored, since the resulting  3N-force model lacks any theoretical backing. Indeed, in the framework of the  coupled-channel approach of Subsection 3.2, Ref.~\refcite{deltuva08} checked the spin-orbit contribution to the 3N force arising from the rho-exchange in the NN to N$\Delta$ transition potential; however, its effect turned out to be far too small to resolve the $\rm A_y$ puzzle in this way. 
\item Ref.~\refcite{deltuva09} used  low-momentum versions of traditional and of $\chi$EFT potentials in the frame work of the $\rm V_{\it low\,k}$ RG and the SRG approaches, in order to study the properties of an additional 3N force needed for resolving the 3N and 4N puzzles. It found that the 3N puzzles seem to require novel longer-ranged isospin-spin properties in the 3N force, whereas the 4N puzzles depend stronger on its short-range parts.
\item
Ref.~\refcite{deltuva10} provides a practical tool for a direct test of parts of the 3N force in their impact on specific observables, i.e., perturbation theory; that tool is technically reliable, but was not followed up yet, since perturbation theory is calculationally for scattering  more demanding than the full direct solution of the 3N and 4N problems.
\end{itemize}

Perhaps it is worth realizing that the practical study of {\em genuine} 3N forces is embedded in a general conceptual question: Besides just adding some sort of 3N force to an assumed 2N-force basis, {\em what is really needed in nuclear forces for a physically reliable description of low-energy nuclear systems?} What can be stripped off from the form of the forces, we got used to, without destroying their predictive power at low energies? This is a challenging question which has been with us for quite some time: \\

\noindent
In the wake of Brueckner theory there was an early attempt to get rid of the explicit construction of potentials. The 2N transition matrix of scattering theory is so close to the Brueckner reaction matrix that its direct construction \cite{sauer03} appears preferable instead of the by-pass through a potential; on-shell the 2N transition matrix is determined by elastic 2N scattering, its pole by the deuteron wave function, half-shell by bremsstrahlung and, completely off-shell, it  directly enters the integral-equation descriptions of the 3N  bound state and 3N scattering. That approach taught us  the most general inverse-scattering theory, a mathematical feast, but it was physicswise not successful, since our theoretical knowledge on the 2N force, i.e., its range and $\pi$-exchange tail could not effectively be incorporated into a direct construction of the 2N transition matrix. The whole approach was also never extended to include 3N forces. \\

\noindent
The same attempt of removing unnecessary aspects from traditional nuclear forces is carried out   \cite{bogner02} by the RG approach of constructing $\rm V_{\it low\,k}$ potentials  and by the SRG approach. In Section 2 I declared those resulting forces {\em induced} ones, technical tools on the way to converged computational results. But in the present context, could one not also view the SRG-evolved potentials as novel {\em genuine} forces, derived from other already tuned {\em genuine} ones and purged from existing unessentials in the latter ones?
\\

\noindent
The same attempt of keeping the dynamic essentials in the nuclear forces is also the basis of $\chi$EFT with its chiral expansion of the interaction processes up to the energetically relevant expansion orders, from which instantaneous potentials for applications are then constructed. The success of $\chi$EFT is due to the fact that it provides a practical answer to the search for the physically important characteristics of nuclear forces, and it provides consistency between the derived {\em genuine} 2N and many-N forces; due to that, it promises to have a rich and successful future; especially, its combination with the RG and SRG approaches offers a fruitful possibility for the practical solution of many nuclear many-body problems. The further conceptual question which remains is: At a given excitation energy of a nuclear system under study, up to which order has the chiral expansion to be pushed and remains manageable, in order to yield converged and therefore theoretically reliable results? Can the $\Delta$-full theory successfully go beyond the $\pi$-production threshold? \\

The nuclear many-body problem is far too complex to be described soon in terms of the underlying QCD degrees of freedom. The intermediary step from QCD to interacting hadronic clusters will therefore be with us for quite some time. 
There are a number of still unresolved problems in that approach. The role of the 3N force will remain in the center of theoretical discussions. 

%%%%%%%%%%%%%%%%%%%%%%%%%%%%%%%%%%%%%%%%%%%%%%%%%%%%%%%%%%%%

\section*{Acknowledgements}

The results, shown in figures, were obtained in a long and fruitful collaboration with A. Deltuva and A.C. Fonseca for which I am thankful. I greatly benefitted from discussions with A. Deltuva, E. Epelbaum, H. Krebs and A. Schwenk on the subject of this review; nevertheless, I am solely responsible for content, style and all possible mistakes of the presentation.

%%%%%%%%%%%%%%%%%%%%%%%%%%%%%%%%%%%%%%%%%%%%%%%%%%%%%%%%%%%%

%%%%%%%%%%%%%%%%%%%%%%%%%%%%%%%%%%%%%%%%%%%%%%%%%%%%%%%%%%%%
\end{document}